\documentstyle[aps]{revtex}

\begin{document}
\title{Understanding entangled spins in QED }
\author{Hai-Jun Wang$^{*}$}
\address{Center for Theoretical Physics and School of Physics, Jilin\\
University, Changchun 130023, China}
\author{W. T. Geng}
\address{Department of Physics, Qingdao University, Qingdao 266071, China and\\
School of Materials Science \& Engineering, University of Science and\\
Technology Beijing, Beijing 100083, China}
\maketitle

\begin{abstract}
The stability of two entangled spins dressed by electrons is studied by
calculating the scattering phase shifts. The interaction between electrons
is interpreted by fully relativistic QED and the screening effect is
described phenomenologically in the Debye exponential form $e^{-\alpha r}$.
Our results show that if the (Einstein-Podolsky-Rosen-) EPR-type states are
kept stable under the interaction of QED, the spatial wave function must be
parity-dependent. The spin-singlet state $s=0$ and the polarized state $%
\frac 1{\sqrt{2}}(\mid +-\rangle -\mid -+\rangle )$ along the z-axis{\bf \ }%
give rise to two different kinds of phase shifts{\bf .} Interestingly, the
interaction between electrons in the spin-singlet pair is found to be
attractive. Such an attraction could be very useful when we extract the
entangled spins from superconductors. A mechanism to filter the entangled
spins is also discussed.


*E-mail address: whj@mail.jlu.edu.cn
\end{abstract}

Quantum entanglement has wide applications in many information processing
protocols, such as quantum teleportation [1], quantum communication and
quantum computation [2], and the development of these quantum technologies
depends strongly on both the properties of the entangled states and our
capability to produce them. The strongly correlated quantum states were
first obtained for photons in 1950 [3], and soon after Bohm expressed the
EPR state with entangled spins [4]. Following these approach, Bell put
forward his famous inequality [5] in 1964, and it was firmly demonstrated
years later in experiments involving only massless photons [6]. Recently,
the entangled spins dressed by relatively isolated electrons in solid-state
materials appear to many researchers as a promising candidate for the
carrier of the EPR states that are required by quantum information
processing [7].

The mechanisms for generating the entanglement of electron spins fall
roughly into two types: one concerns specific systems in solid states [8],
such as generating entangled electron spins via quantum dots or magnetic
impurities, extracting Cooper pairs out of superconductors and etc.. The
other employs special techniques by combining entanglement with the effects
of identity [9, 10] , special relativity [11-14], or electron-electron
interaction within the entangled pair [15, 16, 17, 18]. Among the second
type of mechanisms, of special interests are the space-spin entanglement
transfer [10-13] and generating entanglement by dynamical scattering {\it [}%
17, 18{\it ]}.\ However, serving as information carriers, the entangled
spins are required to be stable enough when used to store and transport
information. The question comes up is what conditions should be satisfied by
the spatial wave function if the spin-singlet pair is to be kept stable
under the interaction between electrons in solids{\it . }

On the other hand, a difficulty encountered in entanglement generation is to
overcome decoherence carsed by the separation of an entangled pair [19] and
the interaction between constituent particles [20]. Therefore, a full
understanding of the electron-electron interaction in the entangled pair is
helpful in the design and control of entanglement. One category of
theoretical mechanisms (thought experiment) to produce electron entanglement
mainly cope with the interaction between electrons through the process of
scattering. Previous studies either employ only an oversimplified
Hamiltonian [15, 16], or analyze the scattering processes using quantum
statistics or some elaborately designed inputs [18] instead of calculating
the amplitude of scattering. In this paper we will describe the intra-pair
spin interaction in a complete form covering all terms related to the spin
operators and calculate the amplitude of scattering. Hence, knowledge gained
in this work is fundamental and will be instructive in constructing any
realistic models of generating entanglement of fermions.

In what follows we focus our investigation on the stability of the EPR state 
$\frac 1{\sqrt{2}}(\mid \uparrow \downarrow \rangle -\mid \downarrow
\uparrow \rangle )$ by calculating the scattering phase shifts. The
interaction between entangled electrons is described by fully relativistic
QED. The method of phase shift has been widely used in many areas such as
particle physics and nuclear physics, but it can only be applied to specific
systems when interactions fall off more rapidly than coulomb potential $1/r$
[21]. In nature, as we know, it is hard to find pure coulomb potential.
Therefore, if we are to examine the interaction between electrons that are
entangled together, it is advisable to include the screening effect. Here,
motivated by the idea of applying Debye theory [22] to the electrolyte,
fluid, and dilute ions [23], we introduce a simple factor $e^{-\alpha r}$($r$
is the distance between the two electrons) to phenomenologically illustrate
this effect. In this picture, the total potential actually falls off more
rapidly than coulomb potential. In solid, while the electrons are moving,
the crystal lattice made up of ions with positive charges and spins around
the electrons will be distorted, yielding a screening effect.

To understand the generation and evolution of the degree of entangled
electron spins, Liu and Chen studied the interaction between the entangled
electrons [16]. The authors analyzed the entanglement of two identical
electrons with an interaction (neither covariant nor screened) interpreted
by the nonrelativistic QED, and showed that the entangled spin-triplet
states can evolve into states bearing no spin-entanglement, whereas the
spin-singlet state remains stable in the scattering process. However, as
shown in the following, our results suggest that to make the spin-singlet
state stable under the electromagnetic interaction, some limitations on the
angular momentum must be added to the spatial wave function.

Here, we take the spin-singlet state of two electrons as initial and final
scattering states to calculate the phase shifts. To make a comparative
study, two methods of directly making total spin $s=0$ (Method A)
[unpolarizing case] and polarizing the orientation of the constituent spins
(Method B) are employed. To avoid confusion, we use $\frac 1{\sqrt{2}}(\mid
\uparrow \downarrow \rangle -\mid \downarrow \uparrow \rangle )$ to denotes
general $s=0$ eigen state appeared in Method A and $\frac 1{\sqrt{2}}(\mid
+-\rangle -\mid -+\rangle )$ the polarized state along the z-axis appeared
in Method B. In quantum mechanics, the above two expressions usually both
denote a spin-singlet state. In QED, however, since we calculate the
unpolarized scattering process by averaging initial spins and summing over
final spins, the results will be dependent on the polarization. So here, we
use these two notations to distinguish two sorts of process (Method A and
B). We will see that although the total spin for both processes are $s=0$,
other features are drastically different. \ Regarding entanglement, the two
states are similar; but for the derivation of scattering amplitude, they
differ from each other.

For the evaluation of the electron-electron scattering phase shift with the
total spin $s=0$ in the initial and final states by means of Method A, we
need the amplitude of the scattering process ${\cal M}_{fi}$, which has the
following form [24]

\begin{eqnarray}
{\cal M}_{fi} &=&(-ie)^2[\overline{u}({\bf p}_3)\gamma _\mu u({\bf p}_1)%
\overline{u}({\bf p}_4)\gamma ^\mu u({\bf p}_2)\frac 1{(p_1-p_3)^2} 
\nonumber  \label{eq.1} \\
&&\ \ \ \ \ -\overline{u}({\bf p}_3)\gamma _\mu u({\bf p}_2)\overline{u}(%
{\bf p}_4)\gamma ^\mu u({\bf p}_1)\frac 1{(p_2-p_3)^2}]  \nonumber
\label{eq.1} \\
\ &=&(-ie)^2[{\cal M}_1\frac 1{(p_1-p_3)^2}-{\cal M}_2\frac 1{(p_2-p_3)^2}%
]\,,  \label{eq.1}
\end{eqnarray}
$u({\bf p})$ is the Dirac spinor defined as $\sqrt{\frac{E+m}{2\,E}}\left( 
\begin{array}{c}
1 \\ 
\frac{{\bf \sigma \cdot \,p}}{E+m}
\end{array}
\right) $, $E=\sqrt{{\bf p}^2+m^2}$. Here the amplitude form is in fact an
operator form, with the wave functions for spins determined by scattering
characteristics{\it . }If the injected states and scattered states are not
polarized, we perform the calculations using Method A, otherwise we use
Method B. The indistinguishability between the entangled electrons is
automatically satisfied by the amplitude. Since this amplitude is covariant,
it is allowable to choose a special reference to simplify the formalism but
at the same time leave the final matrix elements unchanged. Here the
center-of-mass(CoM) reference system is used. Then the electrons' initial
momenta satisfy ${\bf p}_1=-{\bf p}_2={\bf p}$, and the final one has ${\bf p%
}_3=-{\bf p}_4={\bf q}$. For the elastic scattering process, the relation $%
\mid {\bf p}\mid =\mid {\bf q}\mid $ is hold for the momenta. Substituting
the Dirac spinors of the CoM into Eq. (1) leads to the explicit forms of $%
{\cal M}_1$ and ${\cal M}_2$ [25]: 
\begin{eqnarray}
\ {\cal M}_1 &=&\{1+\frac 1{(E+m)^2}[2{\bf q\cdot p+}3i{\bf (q\times p)\cdot
(\sigma }_1+{\bf \sigma }_2{\bf )+q}^2(1-{\bf \sigma }_1\cdot {\bf \sigma }%
_2)+{\bf q\cdot \sigma }_1{\bf q\cdot \sigma }_2  \nonumber \\
&&\ \ \ \ \ \ \ +2{\bf q\cdot p}(1+{\bf \sigma }_1\cdot {\bf \sigma }_2)-%
{\bf p\cdot \sigma }_1{\bf q\cdot \sigma }_2-{\bf q\cdot \sigma }_1{\bf %
p\cdot \sigma }_2+{\bf p}^2(1-{\bf \sigma }_1\cdot {\bf \sigma }_2)+{\bf %
p\cdot \sigma }_1{\bf p\cdot \sigma }_2]  \nonumber \\
&&\ \ \ \ \ \ \ +\frac 1{(E+m)^4}[{\bf q\cdot p+}i{\bf (q\times p)\cdot
\sigma }_1][{\bf q\cdot p+}i{\bf (q\times p)\cdot \sigma }_2]\},
\label{eq.2}
\end{eqnarray}

\begin{eqnarray}
\ {\cal M}_2 &=&\ \{1+\frac 1{(E+m)^2}[-2{\bf q\cdot p-}3i{\bf (q\times
p)\cdot (\sigma _1+\sigma _2)+q}^2(1-{\bf \sigma }_1\cdot {\bf \sigma }_2)+%
{\bf q\cdot \sigma }_1{\bf q\cdot \sigma }_2  \nonumber \\
&&\ \ \ -2{\bf q\cdot p}(1+{\bf \sigma }_1\cdot {\bf \sigma }_2)+{\bf p\cdot
\sigma }_1{\bf q\cdot \sigma }_2+{\bf q\cdot \sigma }_1{\bf p\cdot \sigma }%
_2+{\bf p}^2(1-{\bf \sigma }_1\cdot {\bf \sigma }_2)+{\bf p\cdot \sigma }_1%
{\bf p\cdot \sigma }_2]  \nonumber \\
&&\ \ \ +\frac 1{(E+m)^4}[{\bf q\cdot p+}i{\bf (q\times p)\cdot \sigma }_1][%
{\bf q\cdot p+}i{\bf (q\times p)\cdot \sigma }_2]\}.  \label{eq.3}
\end{eqnarray}
If the $\gamma $-matrix $\gamma _\mu $ in Eq. (1) changes to $\gamma _0$,
the first part of the amplitude reduces to the one that only the interaction
of point charges is included (without the interaction of magnetic moments);
and if furthermore the velocities of the two electrons are very low, the
interaction can be approximately described by the classical Coulomb form $%
\frac 1r$ [26]. In the case $\gamma _\mu $ $\rightarrow \gamma _0$ Eq. (2)
and Eq. (3) reduce respectively to

\begin{eqnarray}
\ {\cal M}_1 &=&\{1+\frac 1{(E+m)^2}[2{\bf q\cdot p+}i{\bf (q\times p)\cdot
(\sigma }_1+{\bf \sigma }_2{\bf )]}  \nonumber \\
&&\ \ \ +\frac 1{(E+m)^4}[{\bf q\cdot p+}i{\bf (q\times p)\cdot \sigma }_1][%
{\bf q\cdot p+}i{\bf (q\times p)\cdot \sigma }_2]\},  \label{eq.4}
\end{eqnarray}
and

\begin{eqnarray}
\ {\cal M}_2 &=&\ \{1+\frac 1{(E+m)^2}[-2{\bf q\cdot p-}i{\bf (q\times
p)\cdot (\sigma }_1+{\bf \sigma }_2{\bf )]}  \nonumber \\
&&\ \ \ +\frac 1{(E+m)^4}[{\bf q\cdot p+}i{\bf (q\times p)\cdot \sigma }_1][%
{\bf q\cdot p+}i{\bf (q\times p)\cdot \sigma }_2]\}.  \label{eq.5}
\end{eqnarray}

The screening effect can be phenomenologically incorporated into the
propagators in Eq. (1) by introducing a factor $e^{-\alpha r}$, where $%
\alpha $ is the inverse of Debye screening length [22] and $r$ the distance
between the two electrons. The factor can be related to the momentum
propagator by applying the Fourier transformation

\begin{equation}
\frac 1{(2\pi )^3}\int \frac{e^{-\alpha r}}re^{i\stackrel{\rightharpoonup }{r%
}\cdot \stackrel{\rightharpoonup }{k}}d^3{\bf k=}\frac 1{k^2-\alpha ^2}\,.
\label{eq.6}
\end{equation}
Obviously, if $\alpha \rightarrow 0$ the propagator will reduce to the
original form. Now we can include the screening effect by using the
propagator $1/(k^2-\alpha ^2)$ instead of $1/k^2$ in Eq. (1). The phenomenal
factor $e^{-\alpha r}$ in the propagator is effective for both scalar and
vector potential, suggesting that the magnetic moments are screened in a
similar way as charges. The screening effect of magnetic moments has been
confirmed by Wilson {\it et al} in studying the Kondo effect [27]. Although
the Kondo model has been studied intensively, its entanglement structure is
still unclear [28]. Up to date no studies of scattering for entangled
screening electrons have been reported.

Now the phase shifts can be calculated under Born approximation, as we did
in a previous study [29] 
\begin{equation}
\delta _l^J=-\frac 12E\,k\,{\cal M}_{fi}^{Jl}(k).  \label{eq7}
\end{equation}
Here $E$ is the total energy of the two-electron system, $k=\mid {\bf p}\mid
=\mid {\bf q}\mid $ is the magnitude of the relative momenta ${\bf p}$ and $%
{\bf q}$, and ${\cal M}_{fi}^{Jl}(k)$, with total angular momentum $J$ and
orbital angular momentum $l$, is the transition amplitude given by

\begin{equation}
{\cal M}_{fi}^{Jl}(k)=\frac 1{(4\pi )^2}\sum\limits_{m,m^{\prime
},m_s,m_s^{\prime }}C_{lm\frac 12m_s}^{JM}C_{lm^{^{\prime }}\frac 12%
m_s^{\prime }}^{JM}\int d\Omega (\stackrel{\wedge }{\bf p})d\Omega (%
\stackrel{\wedge }{\bf q})Y_{lm^{\prime }}^{*}(\stackrel{\wedge }{\bf q}%
)Y_{lm}(\stackrel{\wedge }{\bf p}){\cal M}_{fi}({\bf p},{\bf q}%
;m_s,m_s^{\prime }),  \label{eq8}
\end{equation}
where $C_{lm\frac 12m_s}^{JM}$are the Clebsch-Gordan coefficients, $Y_{l%
\text{ }m}(\stackrel{\wedge }{k})$ the spherical harmonic functions and $%
{\cal M}_{fi}({\bf p},{\bf q};m_s,m_s^{\prime })$ the matrix elements
obtained directly from Eq. (1) by considering the total spins in the initial
and final states: ${\cal M}_{fi}({\bf p},{\bf q};m_s,m_s^{\prime })=\langle
s\,m_{s^{\prime }}\mid {\cal M}_{fi}\mid s\,m_s\rangle $. For special cases
that $s=0$ or $l=0$, i.e. without spin-orbit coupling, Eq.(8) reduces to [30]

\begin{equation}
{\cal M}_{fi}^l(k)=\frac 1{8\pi }\int_{-1}^1dxP_l(x){\cal M}_{fi}({\bf p},%
{\bf q};x),  \label{eq9}
\end{equation}
where $P_l(x)$ is the Legendre polynomial with $x=\cos \theta $, and $\theta 
$ is the angle between ${\bf p}$ and ${\bf q}$.

Now, let us evaluate the scattering phase shifts by means of Method B. The
calculation of polarized amplitudes is analogous to that of a previous study
[24], whereby the spinor $u({\bf p})$ in Eq. (1) is changed to $\sqrt{\frac{%
E+m}{2\,E}}\left( 
\begin{array}{c}
1 \\ 
\frac{{\bf \sigma \cdot \,p}}{E+m}
\end{array}
\right) \chi _\lambda $ to include the spin states $\chi _\lambda $ [ $%
\lambda =1,2$ denote the two spin eigen states $\left( 
\begin{array}{c}
1 \\ 
0
\end{array}
\right) $ and $\left( 
\begin{array}{c}
0 \\ 
1
\end{array}
\right) $]. Here we assume that the momenta of initial electrons lie along
the z-axis, and without losing generality, the spin directions lie along or
opposite to the z-axis. The matrix elements from all possible combinations
of polarized incident and scattered electrons have been listed in Ref. [24].
By imposing a proper transformation with respect to the defined polar angles
of spins on these matrix elements, any other amplitudes with arbitrarily
defined polarized orientation of the initial and final electrons can be
obtained [24]. Expressing the $\{s=0,s_z=0\}$ state with the z-component of
spin angular momentum as $\frac 1{\sqrt{2}}(\mid +-\rangle -\mid -+\rangle )$%
, the scattering matrix element between two $\{s=0,s_z=0\}$ states can be
formally interpreted as $\frac 1{\sqrt{2}}(\langle +-\mid -\langle -+\mid )%
{\cal M}_{fi}\frac 1{\sqrt{2}}(\mid +-\rangle -\mid -+\rangle )$ $=\langle
-+\mid {\cal M}_{fi}\mid -+\rangle -\langle +-\mid {\cal M}_{fi}\mid
-+\rangle $ , where the two terms of RHS are among the aforementioned list
[24]. As there is no ambiguity in defining the related total angular
momentum for partial waves with vanishing total spins, it can be postulated
that Eq. (9) still works in this case if we replace the magnitudes with the
polarized ones.

The signs of the phase shifts can be determined by realizing that while the
interaction between the electrons is interpreted as the classical Coulomb
form [Eq. (4)], the interacting force of the $S$-wave is repulsive, and thus
the phase shifts are negative. The resultant phase shifts of $S$-, $P$-, $D$%
-, and f-wave from Method A are plotted in Fig. 1. The $S$- and $D$-wave
phase shifts obtained with Method B are shown in Fig. 2 [31]. The $P$- and $%
F $-wave shifts obtained with Method B vanish for the reason we will discuss
below. The phase shifts will not change their signs in a wide region of $%
\alpha $ provided that $\alpha $ is lower than the electron mass. Generally,
a larger $\alpha $ corresponds to smaller phase shifts. The dependence is
illustrated in Figures 1 and 2.

Phase shifts obtained with Method A (Fig.1) and B (Fig.2) show much
different features. The phase shifts from two methods for any given partial
wave possess different signs. The sharp difference is presumably due to the
use of different entangled spins in production as the scattering initial and
final states. Although the two entanglement states all satisfy $m_s=0$,
those obtained with Method A seem valid for all directions and those
obtained with Method B only for z direction.{\it \ }The most salient common
feature of the two sets of phase shifts is that they are both
parity-dependent. Fig. 1 shows that the forces of $S$- and $D$-wave are
attractive and those of the $P$- and $F$-wave are strongly repulsive. We
recall that according to the definition of parity, $(-1)^l$, $l$ is the
quantum number in spherical harmonics $Y_{lm}(\theta ,\varphi )$, the $S$-
and $D$-wave are even and the $P$- and $F$-wave are odd. In Fig. 2 we see
that the $S$- and $D$-wave are repulsive, the $P$- and $F$-wave, however,
are all vanishing. The understanding of the parity-dependence may follow the
fact that the even-parity{\bf \ }spatial{\bf \ }function of the two
electrons combined with their $s=0$ antisymmetric spin wave function make up
of the totally perfect antisymmetric wave function required by identical
electrons. Hence, in Fig. 1 the states with the $P$- and $F$-wave spacial
functions are forbidden by the strong repulsive force. In Fig. 2, however,
the forbidden states of $P$- and $F$-wave are automatically removed by the
special polarization in which the spin direction and its relationship to the
momenta (spatial wave function) are defined simultaneously. From
non-relativistic QED it is impossible to obtain the parity-dependence{\bf \ }%
for the spatial wave function.

It's clear that the interaction in Method A is attractive, and that in
Method B is repulsive. The attractive force deserves more attention. Its
order of magnitudes can be evaluated directly from the resultant phase
shifts. With the assumption that the phase shifts are approximately in
proportion to the transmitted momentum $k$ in the region under{\it \ }%
concern, Eq. (7) yields d$\delta _l/$d$k\approx -2M\,\langle \Psi _l\mid
V\mid \Psi _l\rangle \approx -2MV/(197)^3$. For $\alpha =1$, substituting
the electron mass $M=5\times 10^5$eV and d$\delta _l/$d$k\approx 10^{-8}$%
gives $V\approx -10^{-8}$eV. For a smaller $\alpha $, (e.g., $0.001)$, in
low ${\bf k}$ energy region, $V\approx -10^{-4}$eV, on just the same order
of magnitudes as the force of Cooper pair in a superconductor. The results
might be heuristic in the development of the spintronic devices. The
contributions of each term in Eq. (2) and Eq. (3) to the attractive and/or
repulsive forces can be numerically determined in a straightforward way. The
attraction mainly comes from the contribution of ${\bf \sigma }_1\cdot $ $%
{\bf p}_1$ ${\bf \sigma }_2\cdot $ ${\bf p}_2$ $+$ ${\bf \sigma }_1\cdot $ $%
{\bf p}_2$ ${\bf \sigma }_2\cdot $ ${\bf p}_1$, and the repulsion from ${\bf %
\sigma }_1\cdot $ ${\bf \sigma }_2$, ${\bf \sigma }_1\cdot $ ${\bf p}_1$ $%
{\bf \sigma }_2\cdot $ ${\bf p}_1$ and ${\bf \sigma }_1\cdot $ ${\bf p}_2$ $%
{\bf \sigma }_2\cdot $ ${\bf p}_2$. Contributions from the remaining terms
such as the purely coulomb term $\frac 1{k^2}$ are essentially cancelled out
by the subtraction of two terms in Eq.(1).

The characteristic of phase shifts obtained with Method A and B will not
change if we take two-photon processes [16] and radiation corrections into
account. For the two-photon processes, a similar substraction of Eq.(1)
holds too, so the leading contribution of the processes is canceled out. The
remaining terms will not change the signs of the phase shifts because,
multiplied by the square of coupling constant, they are negligible to the
next-leading-order terms of the tree level contribution. So, there is no
need to consider the two-photon processes here. The radiative corrections
can be done by replacing the masses and charges in scattering amplitudes
with the effective ones [32]. They will not change the sign of the
calculated phase shifts for there is no electron's propagator in the
amplitude. To put it another way, the screening effect can be viewed as a
part of the renormalization effect [33].

In light of the above results, we propose a mechanism to produce entangled
pairs of massive spin-1/2 particles. Electron pairs are first injected into
a semiconductor [34] with aforementioned screening\ environment.{\bf \ }In
strong magnetic field spin-orbit coupling can be ignored, and then a
particular relative space-wave-function of pairs, say, $S$-wave or $D$-wave
can be filtered out at the beginning by adding a strong background magnetic
field. Then with a electric field voltage $V$\ the pairs{\bf \ }can be led
to a square potential well with a thin square potential barrier in the
middle which divides the well into two parts (see Fig. 3), the well and
barrier can be formed by normal semiconductor layers [35]. Suppose that only
one pair is allowed in a well for simplicity. The potential step of the well
and the height of the dividing barrier are adjustable just as in practice.
If the interaction is repulsive, the two electrons tend to be separated by
the barrier and the possibility for either of them to tunnel through the
barrier is low; whereas if the interaction is attractive as in the
aforementioned spin-singlet state (Method A), the electrons tend to stay
together in one side and the barrier is easy for them to tunnel through
[36]. Thus the electrical conductivity of the well is very low in the former
case and very high in the latter. In this way the pairs that make the well
very conducting can be filtered out as the maximally entangled spins {\bf (}%
in Method A{\bf )}.

In summary, we have extensively examined the properties of the interacting
entangled electrons in a fully relativistic formalism with two different
methods. Although the techniques used in this work to deal with the
scattering processes are standard and straightforward in QED, the idea to
apply them to electrons in solid-state is very meaningful. Also, the
calculation of tensor forces required in this application is quite
sophisticated and not straightforward. The parity-dependence of the phase
shifts yielded from both approaches suggests that if the entanglement of
spins is kept stable under covariant interaction, the selection of the
spacial wave function will not be arbitrary. Furthermore, we find that the
spin-singlet pair $\frac 1{\sqrt{2}}(\mid \uparrow \downarrow \rangle -\mid
\downarrow \uparrow \rangle )$ and the polarized state $\frac 1{\sqrt{2}}%
(\mid +-\rangle -\mid -+\rangle )$ correspond to two different types of
phase shifts. In the former the electron-electron interaction is attractive
and in the latter it is repulsive. In our thought, it is very important to
demonstrate by deliberate calculations that there is an attraction between
entangled electrons in screening environment. The attraction of like-charged
colloids confined between walls has been observed experimentally [37], and
the screening electron's attraction has since been supposed to exist by some
researchers [38]. However, experimental evidence and theoretical basis for
the latter were still lacking before our work report here. The attraction in
principle can be tested by experiments in solid, and might be helpful in
developing some devices of spintronics.

H. J. W. is grateful to{\it \ }Professor G. M. Zeng and Doctor P. Lyu for
helpful discussions.

\section{Figure Captions}

Fig. 1: The calculated electron-electron scattering phase shifts for {\it S,
P, D, F}-waves from Method A. The dashed, solid, and dotted lines correspond
to $\alpha =0.1,$ $1$ and $10$, respectively.

Fig. 2: The calculated electron-electron scattering phase shifts for {\it S,
P, D, F}-waves from Method B. The dashed, solid, and dotted lines correspond
to $\alpha =0.1,$ $1$ and $10$, respectively.

Fig. 3: A square potential well with a thin square potential barrier in the
middle, and a confined electron pair with {\it S}- or {\it D}-wave
interactions. In spin singlet, the electrons attract each other and the
device containing such wells will display high electrical conductivity;
whereas in polarized case, the electron pairs will contribute little to
conductance.


\begin{references}
\bibitem{}  Here the natural unit $197$eV$\cdot 1$nm $=1$ is used for
convenience of calculation and flat-footed physical meaning. For instance, $%
\alpha =1$ means screened region of a $\sim $200nm.
\bibitem{}  For example see fig.1 in cond-ph/9811338 and 0206323. It is a
customary way to use wells or barriers in quantum mechnics to illustrated
semiconductor layers made up of different materials. Of course there are
many examples in textbooks or papers, we list these references at hand.
\bibitem{}  These results of identical particles associated with the well
with barries are from a exercise problem of quantum mechanics, where I have
replace the bosons with fermions. Results remain similiar.
\end{references}
\end{document}